%% file: Paper.tex
\documentclass[sigconf,screen]{acmart}

\usepackage{array,multirow,etoolbox,graphicx,subcaption,fancyhdr,tabularx,longtable}
\usepackage{float}
\usepackage{graphicx}
\usepackage{amsmath} 
\usepackage{tcolorbox}
\usepackage{booktabs}
\usepackage{mathtools}
\usepackage{xcolor}
\usepackage{hyperref}
\newtcolorbox{TcolorBox}[1]{fonttitle=\bfseries,title=#1}
\usepackage{csquotes}
\usepackage[euler]{textgreek}
\usepackage{microtype}
\usepackage{array,multirow,etoolbox,graphicx,subcaption,fancyhdr,tabularx,longtable}
\usepackage{makecell}
\usepackage{pgfplots}
\usepackage{tablefootnote}
\usepackage{changepage}

\usepackage{xurl} 
\usepackage{tabularx}
\usepackage{caption}
\usepackage{subcaption}
\usepackage{xspace}

\usepackage{enumitem}
\newlist{steps}{enumerate}{1}
\setlist[steps, 1]{label = Step \arabic*:}

\copyrightyear{2022}
\acmYear{2022}
\setcopyright{acmcopyright}
\acmConference[ESEC/FSE '22]{Proceedings of the 30th ACM Joint European Software Engineering Conference and Symposium on the Foundations of Software Engineering}{November 14--18, 2022}{Singapore, Singapore}
\acmBooktitle{Proceedings of the 30th ACM Joint European Software Engineering Conference and Symposium on the Foundations of Software Engineering (ESEC/FSE '22), November 14--18, 2022, Singapore, Singapore}
\acmPrice{15.00}
\acmDOI{10.1145/3540250.3558949}
\acmISBN{978-1-4503-9413-0/22/11}


\input{macros}

\begin{document}
\title{\Title{}}
\input{Abstract}

\input{ccs}

\author{Chandra Maddila}
\email{chandu.maddila@gmail.com}
\affiliation{
  \institution{Microsoft Research}
  \country{USA}
}
\author{Suhas Shanbhogue}
\email{suhas0711@gmail.com}
\affiliation{
  \institution{Microsoft Research}
  \country{India}
}
\author{Apoorva Agrawal}
\email{t-aagraw@microsoft.com}
\affiliation{
  \institution{Microsoft Research}
  \country{India}
}
\author{Thomas Zimmermann}
\email{tzimmer@microsoft.com}
\affiliation{
  \institution{Microsoft Research}
  \country{USA}
}
\author{Chetan Bansal}
\email{chetanb@microsoft.com}
\affiliation{
  \institution{Microsoft Research}
  \country{USA}
}
\author{Nicole Forsgren}
\email{niforsgr@microsoft.com}
\affiliation{
  \institution{Microsoft Research}
  \country{USA}
}
\author{Divyanshu Agrawal}
\email{divagrawal@microsoft.com}
\affiliation{
  \institution{Microsoft Research}
  \country{India}
}
\author{Kim Herzig}
\email{kimh@microsoft.com}
\affiliation{
  \institution{Microsoft}
  \country{USA}
}
\author{Arie van Deursen}
\email{Arie.vanDeursen@tudelft.nl}
\affiliation{
  \institution{Delft University of Technology}
  \country{The Netherlands}
}

% make the title area
\maketitle
\renewcommand{\shortauthors}{C. Maddila et al.}

\input{Introduction}
\input{NalandaGraph}

\input{NalandaIndex}

\input{CodeBook}
\input{Recommender}

\input{Discussion}
\input{RelatedWork}
\input{Conclusion}
%\input{Ack}
%\newpage
%\input{DataAvailability}
\bibliographystyle{ACM-Reference-Format} \balance
\bibliography{Paper}

\end{document}

%% file: Macros.tex
\newcommand{\pr}{pull request}

\newcommand{\ado}{Azure DevOps}
\newcommand{\kg}{socio-technical graph}
\newcommand{\nlg}{Nalanda graph\xspace}

\newcommand{\Title}{Nalanda: A Socio-technical Graph Platform for Building Software Analytics Tools at Enterprise Scale}

\newcommand{\codebookname}{MyNalanda\xspace}
\newcommand{\codebook}{{\codebookname}\xspace}
\newcommand{\originalcodebook}{{Codebook}\xspace}

\newcommand{\DefMacro}[2]{\expandafter\newcommand\csname rmk-#1\endcsname{#2}}
\newcommand{\UseMacro}[1]{\csname rmk-#1\endcsname}

\newcommand{\quotes}[1]{``#1''}

\input{Numbers}

\usepackage{xspace}
\usepackage{dirtytalk}
\usepackage{xcolor}

\definecolor{teal}{RGB}{0,128,128}

\newcommand{\rom}[1]{\uppercase\expandafter{\romannumeral #1\relax}}

\newcommand{\NalandaDocIndex}{Nalanda artifact index\xspace}
\newcommand{\nalanda}{Nalanda artifact and expert recommendation application}

\input{Tables/Numbers}

\newcommand{\expert}{Expert}
\newcommand{\expertLower}{expert}
\newcommand{\companyX}{Microsoft}
\newcommand{\reduceVSpace}{\vspace{-3mm}}
\newcommand{\NalandaGraphPlat}{Nalanda graph system}
\newcommand{\NalandaIndexPlat}{Nalanda index system}

\newcommand{\ArtfactIndexDocuments}{8,018,320}
\newcommand{\ExpertIndexDocuments}{61,428}

%% file: Numbers.tex
\DefMacro{InterviewTotalResponses}{7}
\DefMacro{InterviewTotalResponsesWord}{seven}
\DefMacro{InterviewResponesDevs}{5}
\DefMacro{InterviewResponesDevsWord}{five}
\DefMacro{InterviewResponsesManagers}{2}
\DefMacro{InterviewResponsesManagersWord}{two}
\DefMacro{InterviewResponsesMale}{7}  
\DefMacro{InterviewResponsesFemale}{0}
\DefMacro{InterviewResponsesMaleWord}{seven}  
\DefMacro{InterviewResponsesFemaleWord}{zero}
\DefMacro{InterviewResponse}{\%}

\DefMacro{SurveyTotalParticipants}{2,000}
\DefMacro{SurveyDevParticipants}{1,400}
\DefMacro{SurveyManagerParticipants}{600}
\DefMacro{SurveyTotalResponse}{144}
\DefMacro{SurveyDevResponse}{92}
\DefMacro{SurveyMangerResponse}{44}
\DefMacro{SurveyOOFResponseRate}{10\%}
\DefMacro{SurveyResponseRate}{8\%}
\DefMacro{SurveyAverageResponseTimeInMinutes}{12 minutes}
\DefMacro{SurveyMedianResponseTimeInMinutes}{8.5 minutes}
\newcommand{\NumFavorableResponsesForCodeBook}{74\%} % Apoorva to fill this
\newcommand{\CodeBookDau}{290} % Apoorva to fill this
\newcommand{\CodeBookMau}{590} % Apoorva to fill this

\DefMacro{PrivacyAll}{81.82}
\DefMacro{PrivacyFew}{12.99}
\DefMacro{PrivacyNone}{5.19}

\DefMacro{Algorithmic}{28.5}
\DefMacro{Chronological}{18.18}
\DefMacro{Both}{53.25}

\DefMacro{RunTimeHoursOldBootstrap}{9} %36
\DefMacro{RunTimeHoursNewBootstrap}{28} %40
\DefMacro{RunTimeMinsOldIncremental}{10} %25
\DefMacro{RunTimeMinsNewIncremental}{20} % 30
\DefMacro{RunTimeRecordsNewIncremental}{57K} % 1.25M
\DefMacro{RunTimeRecordsOldIncremental}{10K} % 65K
\DefMacro{RunTimeRecordsOldBootstrap}{1.05M} % 7M
\DefMacro{RunTimeRecordsNewBootstrap}{7.41M} % 150M

\newcommand{\NumRepos}{6,500} %
\newcommand{\NumNodes}{37,410,706} % with auth/rev 37,706,354 , old: 37,375,076 
\newcommand{\NumNodesM}{37M} % with auth/rev 37,706,354 , old: 37,375,076 
\newcommand{\NumEdges}{128,745,590} % {126,950,609
\newcommand{\NumEdgesM}{128M} % {126,950,609
\newcommand{\NumFiles}{14,537,998}
\newcommand{\NumWorkItems}{3,067,754}

%% file: Tables/Numbers.tex
\DefMacro{UserDataSampleSize}{500}

\DefMacro{YoungSmallBaselineAcc}{0.38} % 0.375
\DefMacro{YoungMediumBaselineAcc}{0.36}  % 0.3571
\DefMacro{YoungLargeBaselineAcc}{0.25}
\DefMacro{MediumSmallBaselineAcc}{0.32}
\DefMacro{MediumMediumBaselineAcc}{0.34}  % 0.3415
\DefMacro{MediumLargeBaselineAcc}{0.21}  % 0.2129
\DefMacro{OldLargeBaselineAcc}{0.16}  % 0.1635
\DefMacro{OldMediumBaselineAcc}{0.26}  % 0.2556
\DefMacro{OldSmallBaselineAcc}{0.42}  % 0.4211

\DefMacro{YoungSmallNRRAcc}{0.38}
\DefMacro{YoungMediumNRRAcc}{0.57}
\DefMacro{YoungLargeNRRAcc}{0.50}
\DefMacro{MediumSmallNRRAcc}{0.21}
\DefMacro{MediumMediumNRRAcc}{0.33}
\DefMacro{MediumLargeNRRAcc}{0.39}
\DefMacro{OldLargeNRRAcc}{0.34}
\DefMacro{OldMediumNRRAcc}{0.31}
\DefMacro{OldSmallNRRAcc}{0.21}

\DefMacro{NumberResponseA}{93}
\DefMacro{RateResponseA}{32.40\%}  % 0.32404181
\DefMacro{NumberResponseB}{24}
\DefMacro{RateResponseB}{8.36\%}  % 0.08362369
\DefMacro{NumberResponseC}{170}
\DefMacro{RateResponseC}{59.23\%}  % 0.59233449
\DefMacro{NumberTotalResponse}{287}
\DefMacro{RatePositiveResponse}{67.6\%}
\DefMacro{RateNegativeResponse}{32.4\%}

\DefMacro{NumberFeedback1}{54}
\DefMacro{RateFeedback1}{69.23\%}  % 0.69230769
\DefMacro{NumberFeedback2}{17}
\DefMacro{RateFeedback2}{21.79\%}  % 0.21794872
\DefMacro{NumberFeedback3}{5}
\DefMacro{RateFeedback3}{6.41\%}  % 0.06410256
\DefMacro{NumberFeedback4}{2}
\DefMacro{RateFeedback4}{2.56\%}  % 0.02564103
\DefMacro{NumberFeedbackTotal}{78}
\DefMacro{NumberFeedbackCategory1Combined}{71}
\DefMacro{RateFeedbackCategory1Combined}{91.03\%}

\DefMacro{SmallBaselineAcc}{0.35} 
\DefMacro{MediumBaselineAcc}{0.31}
\DefMacro{LargeBaselineAcc}{0.19}

\DefMacro{SmallNRRAcc}{0.23}
\DefMacro{MediumNRRAcc}{0.36}
\DefMacro{LargeNRRAcc}{0.37}

\DefMacro{Top1RecoveryPct}{1.90}
\DefMacro{Top1RecoveryNum}{97}
\DefMacro{Top3RecoveryPct}{4.50}
\DefMacro{Top3RecoveryNum}{223}
\DefMacro{Top5RecoveryPct}{7.30}
\DefMacro{Top5RecoveryNum}{357}
\DefMacro{Top7RecoveryPct}{8.80}
\DefMacro{Top7RecoveryNum}{432}

%% file: Abstract.tex
\begin{abstract}

Software development is information-dense knowledge work that requires collaboration with other developers and awareness of artifacts such as work items, pull requests, and file changes. With the speed of development increasing, information overload, and information discovery are challenges for people developing and maintaining these systems. Finding information about \emph{similar} code changes and \emph{experts} is difficult for software engineers, especially when they work in large software systems or have just recently joined a project. In this paper, we build a large-scale data platform named Nalanda platform to address the challenges of information overload and discovery. Nalanda contains two subsystems: (1) a large-scale socio-technical graph system, named \emph{\NalandaGraphPlat{}}, and (2) a large-scale index system, named \emph{\NalandaIndexPlat{}} that aims at satisfying the information needs of software developers.

To show the versatility of the Nalanda platform, we built two applications: (1) a software analytics application with a news feed named \codebook{} that has Daily Active Users (DAU) of \CodeBookDau{} and Monthly Active Users (MAU) of \CodeBookMau{}, and (2) a recommendation system for related work items and \pr{}s that accomplished similar tasks (\emph{artifact recommendation}) and a recommendation system for subject matter experts (\emph{expert recommendation}), augmented by the Nalanda socio-technical graph. Initial studies of the two applications found that developers and engineering managers are favorable toward continued use of the news feed application for information discovery. The studies also found that developers agreed that a system like \nalanda{} could reduce the time spent and the number of places needed to visit to find information.

\end{abstract}

%% file: ccs.tex

\ccsdesc[500]{Software and its engineering~Programming teams}

%% file: Introduction.tex
\section{Introduction} 
Building software is a highly collaborative process that requires awareness of the activities by many different stakeholders and interaction with many different artifacts such as files, pull requests, and work items. At the same time, large-scale software development creates lots of data about how people work with each other and with software artifacts. As a consequence, finding information can be hard, especially when software engineers work on large software projects with thousands of files and team members. A lot of times this knowledge about software development activity and expertise is hidden in the form of software development process data and the interaction map between stakeholders and artifacts. This data is hard to mine and represent in a form that allows practitioners to build applications on top of this data. This is primarily due to the scale at which this data is generated and the fact that this data is scattered across disparate data sources and systems. Therefore, it is hard to take full advantage of this data and extract hidden knowledge, without employing a plethora of tailored tools, customized for each source control system and the software development environment.

Socio-technical data that captures social and technical aspects of software development \cite{Sarma2009TesseractIV} is often captured in graph structures.
For example, the Hipikat tool builds a project memory from past activities to support newcomers with software modification tasks \cite{Hipikat}. In 2010, the \originalcodebook framework was introduced with a focus on discovering and exploiting relationships in software organizations to support inter-team coordination \cite{begel2010codebook}. \originalcodebook provided a graph and a query language to support a wide range of applications:\ find the most relevant engineers, find out why a recent change was made, and general awareness of engineering activity \cite{begel2010keeping}. \originalcodebook was built for a single team with 420 developers only. 

With the advent of cloud services, the scale at which software development happens and the volume of data generated during the software development process increased significantly \cite{potvin2016google,ma2019world}. To address the challenges that come with scale, in this paper, we present a large-scale software analytics data platform named Nalanda\footnotemark{} which is built on top of the software development activity data and the artifacts. Nalanda builds a socio-technical graph at \emph{enterprise scale}, with thousands of repositories. Additionally, the \NalandaIndexPlat{}, helps with the search and recommendation of software development artifacts and the experts. Nalanda stores its graph in a native graph database and optimizes heavily to query complex relationships so that software analytics applications can operate directly on the graph via cloud services and a high degree of performance.

\footnotetext{Nalanda is named after an ancient university and knowledge center located in India. It is famous for its huge corpus of scriptures, books, and knowledge repositories.}

\begin{figure}
\centering
\includegraphics[width=0.8\hsize]{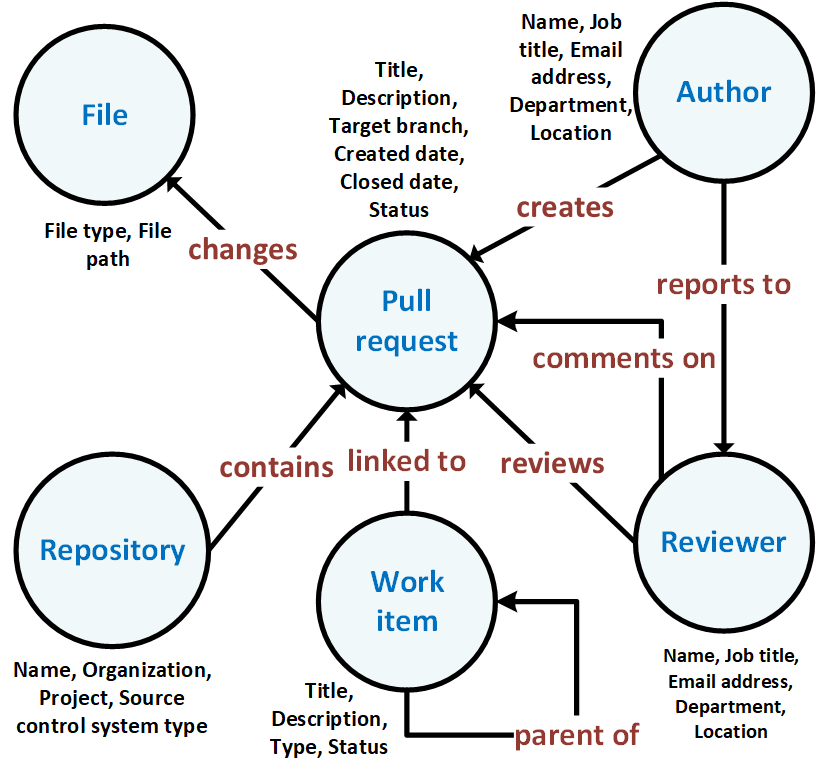}
\caption{Nalanda's Graph Schema}
\label{schema}
\end{figure}

The Nalanda platform, which is a generic and \emph{enterprise scale} software analytics data platform consists of two subsystems: the \NalandaGraphPlat{}, which provides a \emph{large scale} socio-technical graph of software data, and the \NalandaIndexPlat{}, which is an \emph{enterprise scale} index system that can be used to support a wide range of software engineering tasks such as recommendation and search. Nalanda scales to enterprise-scale data from \NumRepos{} repositories. The socio-technical graph has \NumNodes{} nodes and \NumEdges{} edges (The schema of the \nlg{} is shown in Figure~\ref{schema}.) The index system contains 8,079,748 documents.

To show the \emph{versatility} of the Nalanda platform, we describe two tools that have been built on top of the Nalanda platform and \emph{deployed} at \companyX{}: A software analytics news feed application built on top of the \NalandaGraphPlat{} named \codebook{}, and a novel recommendation system (\nalanda{}) leveraging the socio-technical graph for ranking the recommendations.

The goal of this paper is to describe the design, implementation, and deployment of the Nalanda graph system, the index system, and two successful applications (\codebook{} and the \nalanda{}) built and deployed at \companyX{}. We also share details about the extensive analyses and user studies that we conducted to evaluate the perceived usefulness of \codebook{} and \nalanda{} from our deployments at \companyX{}. Additionally, we share insights from building the Nalanda platform, \codebook, and the \nalanda{}.

To that end, we explain the construction of the Nalanda graph system in Section \ref{graph} and the index system in Section \ref{design}. We explain the applications built leveraging these two systems i.e., \codebook in Section \ref{codebook} and the \nalanda{} in Section \ref{recsys}.

%% file: NalandaGraph.tex
\section{Building The Nalanda Graph} \label{graph}
Key challenges in the construction of the Nalanda Graph are scale and consistency.
In this section, we lay out what content we store in the Nalanda graph, 
from which sources we collect the data,
and how we ensure that the graph is kept up to date and consistent as hundreds of thousands of events from thousands of repositories arrive on a daily basis.

\subsection{Nalanda's Graph Schema}
Nodes in the Nalanda graph represent the actors or entities involved in the software development life cycle, 
while the edges represent the relationships that exist between them.

Each node in the \nlg{} has a type associated with it and attributes specific to that node type, as listed in Figure~\ref{schema}.
The central node is the \textsf{Pull Request}, which has incoming edges from \textsf{Author}, \textsf{Reviewer}, \textsf{Work item}, and \textsf{Repository} nodes, and has a outgoing edges to \textsf{File} nodes changed by the pull request.
A developer takes the role of an author when they make source code changes and submit \pr{}s and they assume the role of a reviewer when they perform code reviews. These are represented as user nodes in the \nlg{} with different edge types but are listed here as two different nodes in Figure \ref{schema} for clarity.
For \textsf{File} nodes, different types are distinguished, including source code, configuration, and project files.
Files are edited by the authors via \pr{}s. Files are represented as nodes in the \nlg{} with a second-order relationship established between the user and file nodes via a \textsf{Pull request} node.

Edges in the \nlg{} represent the relationships between various actors and entities. Like nodes, edges can be of different types and can have properties associated with them. An edge is created between an author node and a \pr{} node when a developer creates a \pr{}. Similarly, an edge is established between the reviewer and the \pr{} nodes when a developer is assigned a code review.
%A `contains' edge is created between a repository node and a \pr{} node when a \pr{} is created in a source code repository.
A \textsf{linked to} edge is created when developers link a pull request to a work item, commonly done in Azure DevOps \cite{AzDo} to connect earlier, related, pull requests to new issues.
%A `comments on' edge is created between a reviewer node and a \pr{} node if they add code review comments.
Likewise, a \textsf{parent of} edge is created between two work items if they are linked by the developers with a parent-child relationship in Azure DevOps.
Finally, a \textsf{reports to} edge is created between two user nodes if one of them is the reporting manager of the other.

\iffalse
%\subsection{Augmented \kg{}}
Some of the node types like \textsf{Pull Request}, \textsf{Work item}, and \textsf{File}
have text tokens in them (e.g., in a title, description, path, or name).
These text tokens reflect functional and technical concepts of the system and its application domain.
We capture the relationships between such concepts (represented by word tokens) and the entities and actors, by expanding the \kg{} to have text tokens represented as nodes in the \nlg{} (as shown in Figure \ref{AKG}). We use a simple tokenizer, based on word boundaries, to generate word tokens from the text and remove the stop words \cite{StopWords}. Then, we link the text nodes that appear in a \pr{} text, work item text, and file text to the respective entities. That establishes a second-order relation between text tokens and the user nodes (for authors and reviewers). 
We also establish edges between text nodes based on their co-occurrence in the \pr{} text by using Pointwise Mutual Information (PMI) \cite{manning99foundations}.

\begin{figure}
\centering
\includegraphics[width=0.75\columnwidth]{Figures/Augmented-KG.png}
\caption{The Nalanda Graph with links to text tokens}
\label{AKG}
\end{figure}
\fi

\reduceVSpace{}
\subsection{Data Collection and Graph Construction} \label{NalandaPlatformArch}
\begin{figure}
\centering
\includegraphics[width=0.8\columnwidth]{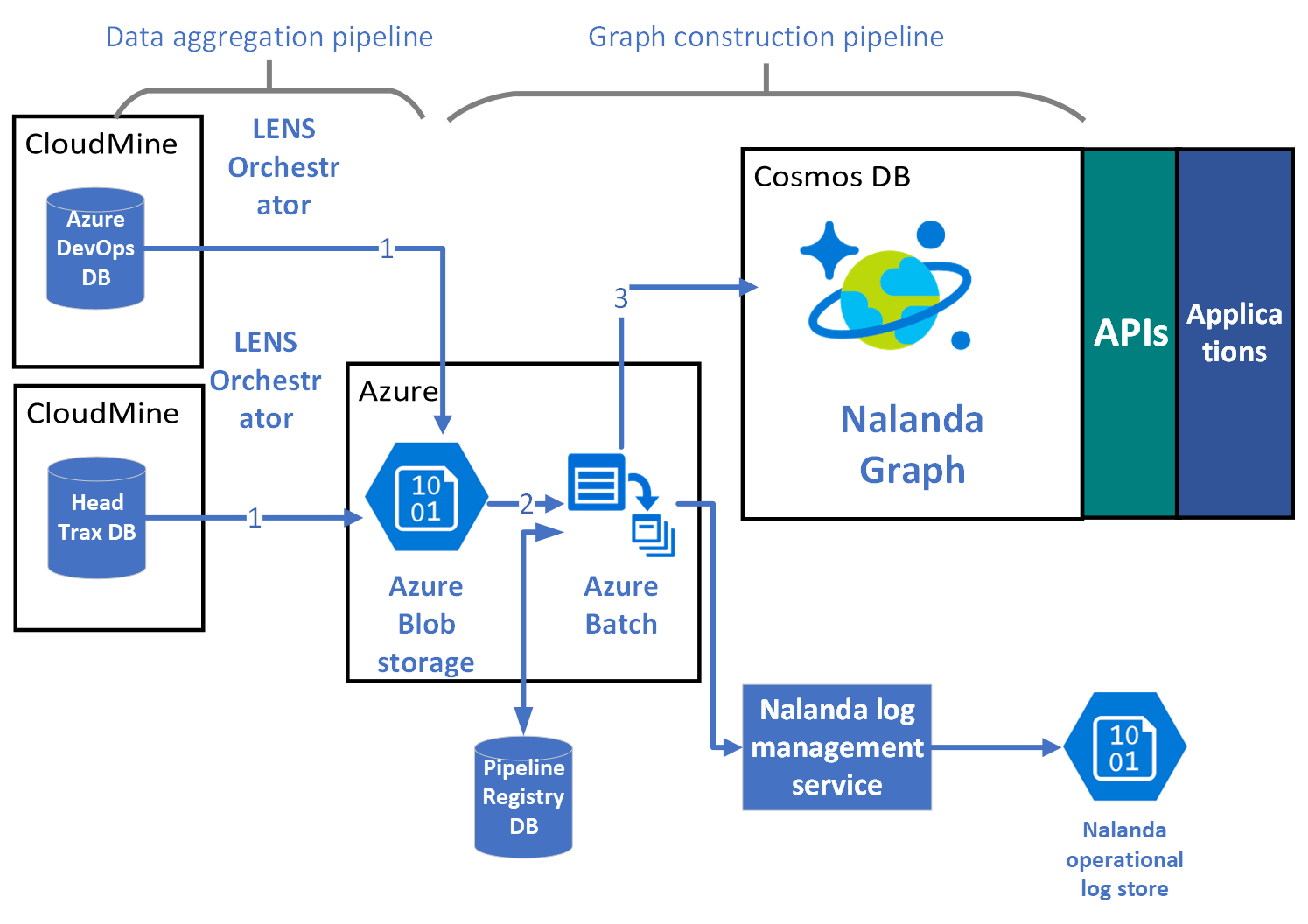}
\caption{Nalanda's data collection and graph construction architecture}
\label{NalandaArch}
\reduceVSpace{}
\end{figure}

The Nalanda platform architecture is shown in Figure~\ref{NalandaArch}. The primary source of data for the \nlg{} is Azure DevOps. Instead of directly crawling the Azure DevOps system for data, we leverage an intermediate data source called CloudMine \cite{czerwonka2013codemine}. The Nalanda platform takes the raw event data from CloudMine and processes it to create the nodes and edges of the \nlg{}. The graph can be queried using the APIs we provide, or 
directly by means of the graph query language Gremlin \cite{Gremlin}.

The platform builds upon Azure \cite{AzService}: key services used include Azure batch, Azure CosmosDB, Azure SQL Server, Azure Blob Storage, and Lens explorer \cite{Lens}.
As shown in Figure~\ref{NalandaArch}, the Nalanda platform is built using two independently operated pipelines: a data aggregation pipeline and a graph construction pipeline, as explained below.

\subsubsection{Nalanda's Data Aggregation Pipeline} \label{DAP}
As indicated as step~1 in Figure \ref{NalandaArch},
the data aggregation pipeline is responsible for fetching data from different data sources (most notably CloudMine) and making it available for the graph construction pipeline to process.

We use Lens Orchestrator \cite{Lens} for orchestration and scheduling purposes. Lens has the ability to connect to multiple data sources and systems and move data around. 
In the aggregator pipeline, Lens first connects to the CloudMine data store (which is hosted on Cosmos \cite{AzService}) and executes Scope scripts to gather data from various data streams, such as the \pr{} stream, the work item stream, the code review stream, etc. Lens saves the aggregated data in the form of CSV files in Cosmos. Then, Lens connects to Azure Blob Storage to temporarily store these CSV files for further processing. This intermediate store is required as CloudMine does not allow any other service (except Lens) to connect to and access the data files for security and compliance reasons. 

Additionally, we use the Lens job scheduling utilities to configure a job in Lens to run once every eight hours to pull the latest data from CloudMine and save it to the Azure Blob Storage.

\subsubsection{Nalanda's Graph Construction Pipeline} \label{GCP}
Once the data is available in Azure Blob Storage, we process it using an Azure batch job to construct the \nlg{} (Steps~2 and~3 in Figure~\ref{NalandaArch}). We use Azure CosmosDB as our graph data store.

When the batch job discovers new data files as generated by the aggregation pipeline,
it updates the pipeline registry with new file information. 
This includes file names, size, timestamp, whether the file contains data from the bootstrap or the incremental stream, file processing status, processing duration, etc. The pipeline registry is a SQL database whose purpose is to serve as a transactional store for the graph construction pipeline. We create one row in the pipeline registry database for every file discovered.
After the new data is downloaded, for each data file the corresponding node and edges are added to, deleted from, or updated in the Nalanda graph. Once all data for a file is read, its registry status is set to ``completed''.

%\subsubsection{Pipeline modes}
The graph construction pipeline operates in two modes: bootstrap mode and incremental mode.

\textit{Bootstrap mode.} This mode helps ingest all of a repository's data, from repository creation time until when it is run. Typically, we run this mode for a repository when it is being onboarded onto the \nlg{} platform for the first time. %When the graph construction pipeline is triggered in bootstrap mode, it also triggers a Lens job to pull data from the original streams of CloudMine. Then, the bootstrap data is copied to a bootstrap folder in Azure Blob Storage, which is subsequently consumed by the graph construction pipeline.

\textit{Incremental mode.} This mode helps keep the data in the \nlg{} updated without needing to read the massive original streams of CloudMine (whose size is in the orders of hundreds of Terabytes), using the incremental streams offered instead (with sizes in the order of tens of Gigabytes). We run the incremental pipeline once every eight hours. 

\input{Tables/PipelineModes}

These separate modes offer the flexibility to bootstrap any new or existing repository data in an independent and asynchronous manner. Furthermore, the separation of bootstrap and incremental pipelines offers substantial performance improvement in terms of run time and resource utilization, as illustrated in Table~\ref{PipelinePerf}.

Refreshing the data by querying the original streams of CloudMine each time the pipeline is run, takes 28 hours for \NumRepos{} repositories.
As the incremental streams are substantially smaller, each incremental job finishes in 20 minutes, yielding an improvement of 98.8\% in pipeline run time, for each pipeline run. Note that for an increase of the number of repositories by a factor of 20, the run time for the bootstrap mode was increased by a factor of 3 only. This is an effect of the careful design and implementation of the data pipeline by massively parallelizing the data processing code and enabling distributed processing on multiple Azure batch nodes. %We also leverage the performance optimizations offered by the Azure CosmosDB to maximize the throughput of our write operations. %\cite{AzureComosDBPerf}.

\subsection{Data consistency and Self-healing} The \nlg{} platform is a distributed system that works with multiple external data sources and large-scale data processing systems, which are prone to introduce data inconsistencies. Data gaps can manifest due to various factors
related to infrastructure and availability of the CloudMine crawlers.

Detecting and remediating data gaps in such a massive distributed system is not a trivial task. We devised a novel self-healing system that detects data gaps and consistency issues proactively and performs self-healing. This helps the pipeline to guarantee data consistency irrespective of the failures manifested in external data sources such as CloudMine. The self-healing system uses the pipeline registry to monitor pipeline states and can switch from incremental to bootstrap mode if this is warranted.

When an incremental pipeline is run, the timestamp of the oldest record to be processed is compared with the timestamp of the last successful pipeline run.
If the difference between these timestamps is bigger than three days (the period for which incremental streams hold their data), this means a data gap has occurred. To address this, the bootstrap mode is triggered for the repositories involved, and ongoing incremental pipelines are halted. Furthermore, the pipeline registry is updated to indicate that bootstrapping is taking place, thereby locking new incremental jobs.
%\reduceVSpace{}
\subsection{Scale} \label{stg-sclae}
The \nlg{} has been designed to accommodate thousands of repositories.
At the time of writing, it holds the software development activity data from \NumRepos\ repositories at Microsoft. We ingest data starting from January 1, 2019, or from more recent repositories when their first \pr{} is created.

To keep its graph up to date, the Nalanda platform processes 500,000 events per day. These events include new \pr{}s, updates or commits on those \pr{}s, \pr{} state changes, code review assignments, and code review comments.
At the time of writing, the \nlg{} contains 37 million nodes and 128 million edges as detailed in Table~\ref{tab:NodeEdgeDistribution}.

%\arie{Why are the "repository" nodes not included in the table? Or are there no such nodes? Explain?
%Are the "text" nodes single tokens, or larger pieces of text? How do these relate to the augmented Nalanda graph?
%}

\begin{table}
\centering
\caption{Nalanda node and edge types and their prevalence}
% \begin{tabular}{llr}
% \toprule
% Element type & Label & Count \\
% \midrule
% Node  & \pr{} & 7,337,036  \\
% Node  & work item & 2,972,785  \\
% Node  & file & 14,533,602  \\
% Node  & author & 131,578  \\
% Node  & reviewer & 295,648  \\
% Node  & text & 12,104,427  \\ 
% \midrule
% \textbf{Total} & & \textbf{\NumNodes{}} \\
% \midrule
% Edge  & creates & 7,337,125  \\
% Edge  & reviews & 38,300,628  \\
% Edge  & contains & 7,337,036  \\
% Edge  & changes & 65,691,018  \\
% Edge  & parent of & 818,738  \\
% Edge  & linked to & 6,887,797  \\
% Edge  & comments on & 578,267  \\
% \midrule
% \textbf{Total} & & \textbf{\NumEdges{}} \\
% \bottomrule
% \end{tabular}
% \label{tab:NodeEdgeDistribution}
% \bigskip
\vspace{-0.5\baselineskip}
\small
\begin{tabular}{lr}
\toprule
Node type & Count \\
\midrule
file & \NumFiles{}  \\
%author & 131,578  \\
%reviewer & 295,648  \\
text & 12,104,427  \\ 
\pr{} & 7,568,949  \\
work item & \NumWorkItems{}  \\

user & 131,578\\
repository & \NumRepos\\
& \\
\midrule
\textbf{Total nodes} & \textbf{\NumNodes{}} \\
\bottomrule
\end{tabular}\hspace{0.6cm}%\hfill
\begin{tabular}{lr}
\toprule
Edge type & Count \\
\midrule
changes & 65,706,621  \\
reviews & 39,447,635  \\
creates & 7,569,086  \\
contains & 7,337,036  \\
linked to & 7,094,597  \\
parent of & 843,728  \\
comments on & 746,887  \\
\midrule
\textbf{Total edges} & \textbf{\NumEdges{}} \\
\bottomrule
\end{tabular}
\label{tab:NodeEdgeDistribution}
\end{table}

%% file: Tables/PipelineModes.tex
\begin{table}
\centering
\caption{Comparison of pipeline run time and number of records with the increase in number of repositories}
\small
\begin{tabular}{*5l}
\toprule
Mode &  \multicolumn{2}{c}{Run time} & \multicolumn{2}{c}{\thead{\# Records to process}}\\
\midrule
{}   & 350 repos  & 6500 repos    & 350 repos   & 6500 repos\\
\midrule
Bootstrap   &  \UseMacro{RunTimeHoursOldBootstrap} hrs & \UseMacro{RunTimeHoursNewBootstrap} hrs   & \UseMacro{RunTimeRecordsOldBootstrap}  & \UseMacro{RunTimeRecordsNewBootstrap}\\
Incremental   &  \UseMacro{RunTimeMinsOldIncremental} min & \UseMacro{RunTimeMinsNewIncremental} min   & \UseMacro{RunTimeRecordsOldIncremental}  & \UseMacro{RunTimeRecordsNewIncremental} \\
\bottomrule
\end{tabular}
\label{PipelinePerf}
\reduceVSpace{}
\end{table}

%% file: NalandaIndex.tex
\section{Indexing Nalanda for Information Retrieval} \label{design}

Many nodes in the Nalanda Graph contain text. To facilitate \emph{search} over such text at the Nalanda scale, we need to create appropriate indexes.
The actual indices needed may depend on the specific applications built on top of the Nalanda graph.
In this section, we discuss the indices we create and how we ensure they remain up to date at scale.

For every search, we use the BM25 algorithm \cite{RobertsonWHGL92} for determining text similarity between a query and documents (\pr{}s, work items, ...) and ranking the results. BM25 is a bag-of-words model developed based on the probabilistic retrieval framework \cite{article-prob-ret}. For a given query Q containing keywords q\textsubscript{1},...,q\textsubscript{n}, the BM25 score for a document D is:

\begin{equation} \label{eq:bm25} \text{score}(D,Q) = \sum_{i=1}^{n} \text{IDF}(q_i) \cdot \frac{f(q_i, D) \cdot (k_1 + 1)}{f(q_i, D) + k_1 \cdot (1 - b + b \cdot \frac{|D|}{\text{avgdl}})} \end{equation}
where $f(q_i, D)$ is $q_i$'s term frequency  in the document $D$, IDF($q_i$) is $q_i$'s inverse document frequency, $|D|$ is the length of the document $D$ in words, and avgdl is the average document length in the text collection from which documents are drawn. $k_1$ and $b$ are free parameters.
We use standard recommended values ($b=0.75$, $k_1 = 1.2$)
for these constants \cite{schutze2008introduction}.

\subsection{The Nalanda Artifact Index} \label{docIndex}

The \NalandaDocIndex facilitates search through pull requests and work items. We index the metadata, titles, and descriptions of the artifacts (\pr{}s and work items). The metadata consists of elementary properties, namely project name, repository name, and organization name.

%An important point to take into consideration while building the artifact index is to strike a good balance between the number of attributes of work items and \pr{}s that need to be indexed versus the accuracy of the search results. The more attributes we index the bigger the index will become and impact the query performance. We performed a comparative study to derive the best combination of the attributes. The results of the study are detailed in Section \ref{eval}.

%The corpus used by the \NalandaDocIndex{} comprises two types of documents: \pr{}s and work items. We index the metadata, titles, and descriptions of the artifacts (\pr{}s and work items). The metadata consists of elementary properties, namely project name, repository name, and organization name. 

The \NalandaDocIndex can be used to find relevant pull requests given a work item, feature, technical, or functional concept.
Furthermore, with this index 
completed \pr{}s and work items can be used as a template and inspiration to solve similar problems. They provide code samples, expose code review comments, and help as informal documentation to learn the best practices. Additionally, they help understand the team or project-specific processes involved in getting such \pr{}s completed.

%\subsection{Background: Ranking using BM25}

%The artifact index is lightweight as it is built on top of the corpus that contains the text tokens extracted from the titles and descriptions of \pr{}s and work items.

%Unlike code search \cite{CodeSearchGH}, where the primary intent is to index the source code files in a project, we index the software development activity data (\pr{}s and work items). The Nalanda recommendation system differs from traditional code search in two ways: 1) Code search is helpful to find the files in which a function resides, if one knows the exact name of the function and the project in which it resides. For example, one can search for a function name, such as \texttt{ImapProtocolHandler}, and find its implementation. However, it does not support search queries like `how to implement IMAP protocol handler in Orion mail client'. 2) Code search helps find where a class or a function resides, but it does not necessarily help find the details about the process building functionality. This includes the code changes that need to be made, the dependencies that need to be taken care of, test suits that need to be run, processes that need to be followed to push the changes to deployment, etc. Additionally, code search indices consume a lot of space and grow big with time. 

%For the concept search task, Q is the input query entered by the user and the document corpus remains the same. This helps us in reusing the same document corpus (and index) for both the tasks.

\subsection{The Nalanda Expert Index} \label{peopleIndex}
The Nalanda expert index is built to map subject matter experts (SMEs) to technical and functional skills. 
\expert{}s can be of two types: functional and technical.
Functional \expertLower{}s have expertise in specific functionality of a software product or service, such as the query optimizer in an RDBMS product or the ranker in a search engine product.
Technical \expertLower{}s have expertise with a technology concept such as socket programming in Java.

The Nalanda \expertLower{} index relies on a collaborative software development platform like \ado{} to mine and associate expertise with people. We match pull requests and work items (as also used for the artifact index) to their authors and contributors. The process of building the Nalanda \expertLower{} index consists of the following steps:
\begin{enumerate}
    \item Find all  \pr{}s and work items completed by a developer and extract key phrases from them. The key phrases include tokens in \pr{} minus English stopwords \cite{StopWords}.
    \item Create a document using the data from step 1, which is the representation of a developer's skills.
    \item Repeat steps 1-2 for every developer and builds the corpus for the \expertLower{} index
\end{enumerate}

Similar to the \NalandaDocIndex{}, we use the BM25 algorithm for querying the \expertLower{} index corpus.
The intuition is that if a developer makes frequent code changes related to a topic (functional or technical), they must be knowledgeable in that topic area. We represent the frequency of a topic in a developer's activity as term frequency in the BM25 index. Therefore, the more a topic appears in the document corpus constructed for that developer, the more weight that topic is given.

We leverage the Nalanda socio-technical graph for re-ranking the search results returned by the artifact and expert indices. A detailed analysis of the impact of employing the Nalanda graph in refining the search results is discussed in Section \ref{eval}.

\subsection{Scale}
Building indices such as the artifact and expert index at scale is an expensive operation. We carefully crafted the system design to make the Nalanda index creation and refresh pipelines robust and tolerant to failures (details about implementation are explained in Section \ref{Impl}). The Nalanda search system has been built as a cloud-native service. This enables us to scale out the system horizontally with the increase in data and query volume. This also helps us in meeting high uptime Service Level Agreements (SLAs) requirements to move to production. Currently, the artifact index contains 8,018,320 documents and the \expertLower{} index contains 61,428 documents. 

We ingest data from \NumRepos{} repositories. Our index data refresh pipeline, which runs once every week, completes in 65 minutes on average. The graph data refresh pipeline, which runs every 8 hours and finishes in 18 minutes. We optimized the API service to return the response in 1.7 seconds in accordance to the SLA requirements.

%% file: CodeBook.tex
\section{Nalanda Applications (I): The \codebook Portal} \label{codebook}

The Nalanda graph and indexing platform can be used to build many applications to 
support software development teams and organizations in their daily work.
The first application we discuss is \codebook,
an online news feed in production at Microsoft, in which developers and managers alike can monitor ongoing software development activities.

\subsection{\codebook Motivation}
The motivation behind \codebook{} is that
it is common practice for developers to work on multiple work items or \pr{}s at once. It is also common practice for developers at Microsoft to work on multiple source code repositories simultaneously. Microsoft does not have many large mono-repositories, but a lot of small or medium sized repositories. Keeping track of one's work items in their repository or across multiple repositories is a difficult and time consuming task. %Azure DevOps provides a query editor \cite{QueryEditor} utility, but it is complex to use and it does not support multi-repository scenarios. 
Moreover, these tools operate in a {workitem-centric} fashion, i.e., the primary goal of the tool is to search for and find a work item one is interested in.

By contrast, \codebook{} is a \emph{developer-centric} news feed. Upon login, \codebook{} shows the activity (\pr{}, work item, code review) of a developer, from multiple repositories, in their homepage. Additionally, \codebook{} enables developers to discover what their teammates and other collaborators are working on without the hassle of going to different Azure DevOps repositories. %(if they know the repository names and locations) to fire multiple complex queries in the query editor.

\begin{figure}
\centering
\includegraphics[width=1\columnwidth]{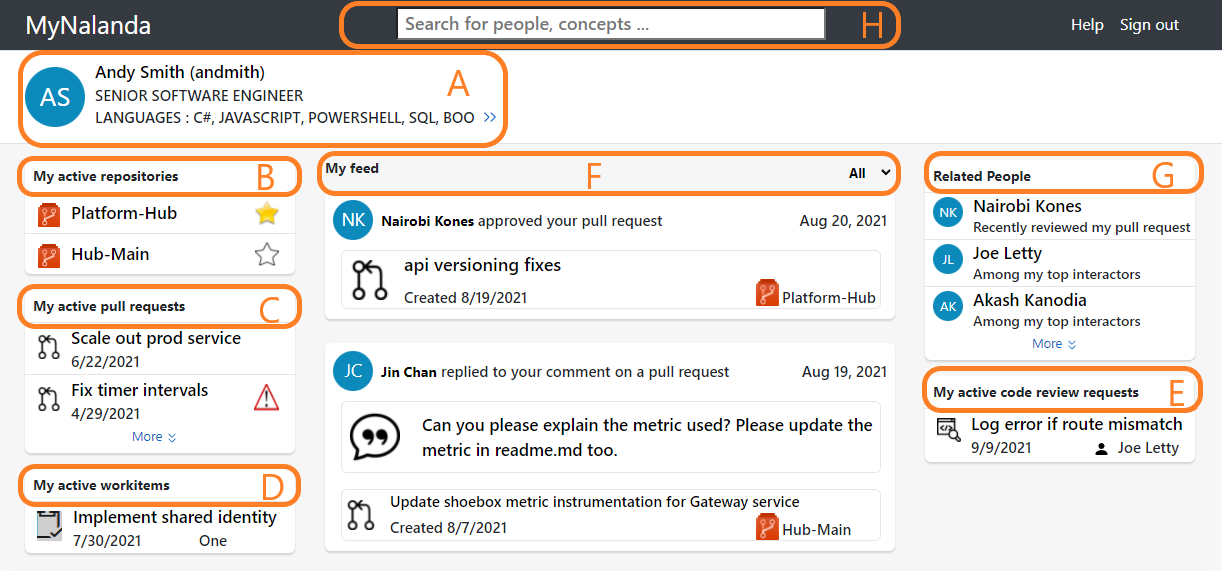}
\caption{The \codebook{} homepage}
\label{codebooksample}
%\reduceVSpace{}
\end{figure}

\subsection{The \codebook Homepage} \label{MyNalandaHomepage}
For a user, the central hub in \codebook is their \emph{homepage}.
An example of such a \codebook{} homepage is shown in Figure \ref{codebooksample}.
Its information is organized in the following sections:

\textbf{News feed}: The centralized news feed (`F' in Figure~\ref{codebooksample}) is located in the middle of the page. The news feed shows events such as updates in \pr{}s, code review comments, and \pr{} status changes from all the repositories a developer works in. For managers, the news feed provides updates from their reports' activity.
%Users have the ability to order the events in the news feed chronologically or by repository. Users can also apply filters on the repositories from which they would like to consume events.

\textbf{User details}: This section (`A' in Figure~\ref{codebooksample}) provides details about developers, such as name, email address, job title, and their expertise (extracted from their software development activity data). This helps in facilitating easy discovery of developers' skills and their current projects.

\textbf{Active items}: There are separate sections for active repositories, \pr{}s, work items, and code review requests (`B'-`E' in Figure~\ref{codebooksample}). Users of \codebook can prioritize the discovery of updates from these items by \emph{following} or \emph{unfollowing} them. %These items (and their updates) are shown at the top of the news feed. Users can also unfollow an item if they would like to stop receiving updates about such repositories.

\textbf{Related people}: This section (`G' in Figure~\ref{codebooksample}) visualizes who a developer collaborates with and how local software development communities are formed. A developer's collaborators include others who work together with the developer on a coding task or work item, or who are either being reviewed by that developer, or who are involved in reviewing a developers' pull request.

\textbf{Search box}: The search box (`H' in Figure~\ref{codebooksample}) can be used to find developers and discover their activity. It also can help with searching for technical and functional concepts by leveraging the Nalanda artifact and expert indices (as explained in Section \ref{design}).

All elements in \codebook{}, such as \pr{}s, work items, people, and repositories have embedded URLs which take them to the corresponding item in Azure DevOps. This makes it easy for developers to navigate between \codebook{} and Azure DevOps.
%Users can see their own homepage, as well as the homepages of other developers, making it easy to see what others are working on.

Additionally, \codebook facilitates integration of other machine learning recommenders due to its extensible architecture. For example, overdue \pr{} are indicated with a subtle warning icon in the active \pr{}s section. This is powered by the Nudge machine learning models \cite{chandraNudgepaper}.

\subsection{\codebook Usage}

\codebook leverages the graph representation of the data (the \nlg{}) and its schema design to navigate efficiently through complex relationships and find the content presented in various sections.
As a result, the \codebook homepage including the news feed and the other sections loads in less than a second.
As quoted by one of the \codebook users \textit{``It is simple, blazing fast to load, adapts to screen size''}. 

Based on organic growth alone, \codebook{} reached 290 Daily Active Users (DAU), and 590 MonthlyActive Users (MAU), in the first six months of deployment of the Beta version at Microsoft.

\subsection{\codebook Evaluation: Perceived Usefulness} \label{UserStudy}
To evaluate how developers and engineering managers perceive the usefulness of \codebook, we follow a mixed method research design involving interviews and surveys.
With increasing sources of development-related information available, there remains an open question about how they want to access and integrate information about their own development activities and the development work done by their peers, and if current platforms are adequate.
Through this evaluation we assess how \codebook matches the corresponding information needs.
\subsubsection{Evaluation Setup} 
\paragraph{Semi-structured Interviews.}
We conducted interviews to investigate information discovery and overload, and if users might use an interface like \codebook{}. Participants included \UseMacro{InterviewResponesDevsWord} developers and \UseMacro{InterviewResponsesManagersWord} engineering managers; \UseMacro{InterviewResponsesMaleWord} participants were men and \UseMacro{InterviewResponsesFemaleWord} were women. Semi-structured interviews were conducted remotely, and ranged from 30-45 minutes. Interview topics included interest in accessing information about their own and peers' development activity, information overload and how they typically get information (in both in-office and work-from-home contexts). We then showed a deployed version of \codebook{} and asked for reactions including if they would like it, what information they would find useful, and where they would want to see it. If the interviewee was an engineering manager, they were also asked what information they would be interested in seeing related to their team's work. 

Immediately following the interviews, notes were taken by the interviewer to augment the transcription and interviews were coded for emergent themes. Following each subsequent interview, themes were revisited to see if any codes should be combined or separated. Once no new themes emerged (i.e., theoretical saturation), we concluded our interview phase. After \UseMacro{InterviewTotalResponsesWord} interviews, themes remained consistent. Finally, we reviewed notable excerpts from all interviews and organized the themes by topic. 

\reduceVSpace{}
\paragraph{Surveys.} Following interviews, we conducted surveys to validate and quantify the themes that emerged. Survey participants included full-time employees who were developers or engineering managers. %While their work spanned various areas across Microsoft, all had contributed to at least one of the \NumRepos{} repositories included in the Nalanda Graph, which allowed participants to access their \codebook{} homepage during the survey.

We designed our survey based on themes that emerged during our interviews, resulting in a 19-item instrument that took a median of \UseMacro{SurveyMedianResponseTimeInMinutes} to complete. Participants were shown their \codebook{} newsfeed (with their own development activity) and given the survey. Topics included demographics, usefulness of information included in their \codebook{} feed, information pain points and privacy concerns, and current and anticipated work location (e.g., office or work-from-home). We included items asking about usefulness of \codebook{} information, and preferences for possible features (based on jobs-to-be-done). We asked where respondents would like to see \codebook{} integrated (if at all), and their comfort level in sharing their development activity through something like \codebook{}. The full survey instrument is available online at research.microsoft.com~\cite{maddila2021nalanda}.

The survey was sent to \UseMacro{SurveyTotalParticipants} people in total (\UseMacro{SurveyDevParticipants} developers and \UseMacro{SurveyManagerParticipants} engineering managers) %in the United States and India ,
with \UseMacro{SurveyTotalResponse} responses (\UseMacro{SurveyDevResponse} developers and \UseMacro{SurveyMangerResponse} engineering managers), resulting in an \UseMacro{SurveyResponseRate} response rate after considering the \UseMacro{SurveyOOFResponseRate} out of office responses. Our low response rate could be due to the fact that the survey is not a trivial one to fill in (cognitively and time it takes to complete the survey) and/or because we did not offer any incentives for participation. Of those respondents, 92 (67.65\%) were developers (including software development engineers, senior software engineers, etc.) and 44 (32.35\%) were engineering managers (including software engineering manager, software engineering lead, etc.). Our respondents included five women (6.17\%), 73 men (90.12\%), one who preferred to self-describe (1.32\%), and two who preferred not to answer (2.63\%). They reported an average of 10.15 years working at Microsoft, ranging from 0.6 to 29.9 years (standard deviation 7.13).  

%\reduceVSpace{}
\subsubsection{Results of the Study}
We first present findings from our semi structured interviews, with each noted as developer (D) or engineering manager (EM). We then present our survey results. 

\paragraph{Interviews.} Participants expressed two themes related to information needs: integration and overload. Information integration was echoed by many interviewees; we define this as having development-related information for self and others integrated into a single, easy-to-access interface. P1 (EM) and P2 (D) discussed easily \textit{linking} design docs and associated artifacts like pull requests. P1 (EM) discussed the usefulness of graphic summaries for their teams' development activity across time periods; this reflects consolidation via visualization. P2 (D) called out that the ability to see detailed information about peers' work is helpful when coordinating work, and is not readily available. %The importance of information integration is echoed in prior studies investigating information characteristics of technical tools used to build systems \cite{forsgren2016integrated}.

%Timely information was another theme that emerged in our interviews. P6 (D) stated, "The overdue information in the active \pr{} section is really helpful for new engineers."  -- timeliness might not matter much given our new framing

P5 (D) spoke about viewing the information in different ways to deal with information overload and ensure they were not missing things: using a `Most recent' view for chronologically-ordered information, an algorithmic `Relevance' view to combine information across teams, and team-only view to further filter. This speaks to strategies they use to ensure keep up on development-related information both within and across teams they collaborate with. Similar sentiments were expressed by other interviewees.

\paragraph{Surveys.} We asked participants to rate how useful each feature of \codebook{} was on a five-point Likert-type scale ranging from 1= Not at all useful to 5= Extremely useful. All items were optional; of those who took the survey, 85 answered questions about \codebook{} features (63 developers and 22 engineering managers). This allowed us to capture participant reactions to a real integrated information platform instead of a hypothetical one.

Table~\ref{tab:newsfeedFeedbackwithBars} lists the accumulated percentages of \quotes{Extremely useful} (Likert=5) and \quotes{Very useful} (Likert = 4) for each feature. Here, we see that \textit{Active pull requests}\textit{} (55.6\% and 54.5\% among developers and engineering managers, respectively) and \textit{Active code review requests} (49.2\% and 45.5\% among developers and engineering managers, respectively) are the highest rated features, with \textit{User details} (27.0\% and 38.2\% among developers and engineering managers, respectively) and \textit{Related People} (23.8\% and 9.1\% among developers and engineering managers, respectively) rated the lowest. %The biggest difference between developers and engineering managers is seen in the \textit{Feed} (34.9\% and 9.1\% among developers and engineering managers, respectively). That is because, very few engineering managers actively create \pr{}s and naturally do not find a news feed with their \pr{} activity data enticing.

Based on this, we conclude that the active items (pull requests, code review requests, repositories, and work items) are the most valued features of \codebook, and that user details are primarily of interest to engineering managers.

%trying more versions for the above table
\input{Tables/NewsfeedFeedbackv1Barv1}

%\input{Tables/NewsFeedback1BarDelta}

%To evaluate information overload, we asked \quotes{How challenging is it for you to currently discover and manage development activity that is relevant for you?} Among our survey respondents, 53\% indicate it is a minor concern and 47\% indicate it is a major concern. If given an opportunity to skip some information to help deal with information overload, respondents indicated \quotes{comments and replies... there is so many of these it is extremely noisy} and \quotes{I would like to see current state... on the level of work items and pull requests.} The Nalanda platform provides for additional applications that can address these challenges for users.

%% file: Tables/NewsfeedFeedbackv1Barv1.tex
\pgfplotsset{width=10cm,compat=1.7}

\begin{table}[t]\centering

\footnotesize
\newlength{\myboxheight}
\settoheight{\myboxheight}{1234567890\%}

\def\mybarchart#1{
\resizebox {#1} {\myboxheight} {%
\begin{tikzpicture}[]
\definecolor{clr1}{RGB}{99,99,99}
\definecolor{clr2}{RGB}{240,240,240}
\begin{axis}[
axis background/.style={fill=gray!10, draw=gray!50},
axis line style={draw=none},
tick style={draw=none},
ytick=\empty,
xtick=\empty,
ymin=0, ymax=1, %1 % this is 0.70 here, the other table is 0.60
xmin=0, xmax=1] %1
\addplot [
ybar interval=.25, %0.5
fill=black,
draw=none,
]
coordinates {(1,1) (1,1)};%1,1 1,1
\addplot [
ybar interval=0.5,%.5
fill=black,
draw=none,
]
coordinates {(1,1) (0,1)};%
\end{axis}%
\end{tikzpicture}%
}%
}

\caption{\codebook survey Feedback}
\label{tab:newsfeedFeedbackwithBars}

%\resizebox{\textwidth}{!}{%
\small
\begin{tabular}{p{2.75cm}|p{2cm}|l|l}
\toprule
Feature & \multicolumn{1}{l|}{\thead{Cumulative\\ (n=85)}}  & \thead{Developers\\(n=63)} & \thead{Managers\\(n=22)} \\
%& \thead{Delta(Developers-\\Engineering\\Managers)} \\
%\cmidrule(r){1-1}\cmidrule(lr){2-2}\cmidrule(l){3-6}
\midrule
Active pull requests (C) & \mybarchart{27.65pt} 55.3\% & 55.6\% & 54.5\%  \\
Active code review requests (E) & \mybarchart{24.2pt} 48.2\%  & 49.2\% & 45.5\%  \\
%\midrule
Active repositories (B) & \mybarchart{22.35pt} 44.7\%  & 46.0\% & 40.9\% \\
%\midrule
Active work items (D) & \mybarchart{17.55pt} 34.1\%  & 34.9\% & 31.8\% \\
Feed (F) & \mybarchart{14.1pt} 28.2\% & 34.9\% & 29.1\% \\
User Details (A) & \mybarchart{12.35pt} 24.7\% & 27\% & 38.2\% \\
Related people (G) & \mybarchart{10pt} 20\% & 23.8\% & 9.1\% \\
\bottomrule
\end{tabular}
\end{table}

%% file: Recommender.tex
\section{Nalanda Applications (II): Artifact and Expert Recommender} \label{recsys}
\begin{figure}
\centering
\includegraphics[width=1\columnwidth]{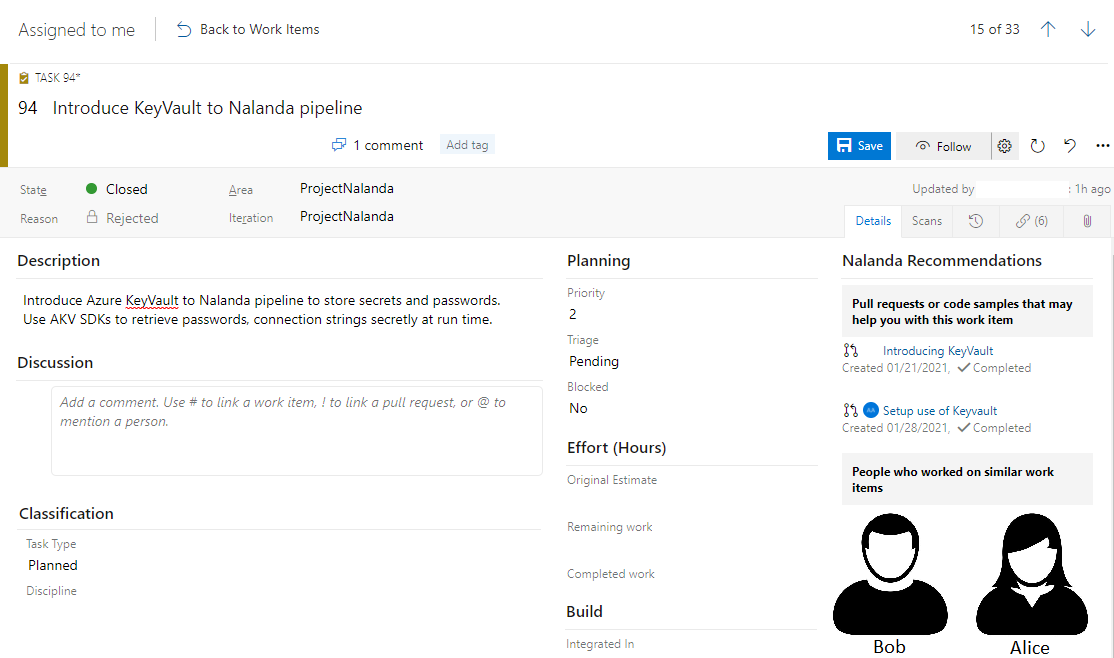}
\caption{Nalanda recommendations in \ado{}}
\label{WIRecommendation}
\end{figure}

%\subsection{Artifact and Expert Recommender motivation}
When a developer is working on a work item or feature, finding the relevant \pr{}s and code samples is one of the biggest pain points for developers ~\cite{maddila2021nalanda}. The intensity of the problem multiplies in large organizations with teams working on multiple source code repositories. Similarly, finding functional and technical experts in large organizations is a difficult task. A lot of times, this involves the developers to mine git history, going through wiki pages, design documents, etc. 

The \nalanda{} implements a recommendation plug-in for Azure DevOps (AzDO) \cite{AzDo}. When a new work item is assigned to a developer, the plug-in triggers an API call to the Nalanda search API. The client passes the necessary input parameters, such as the search query (work item title and description), the work item owner, and the repository metadata. Upon receiving the results (artifact and \expertLower{} recommendations), the client add recommendations to the work item page (in Azure DevOps). A sample work item recommendation page from live deployment (Beta version) is shown in Figure~\ref{WIRecommendation}. Besides, through its easy-to-use APIs, the \nalanda{} system powers other applications such as the \emph{search box} in MyNalanda (as explained in Section \ref{MyNalandaHomepage}).

%\subsection{The Nalanda Expert and Artifact Recommender}

%\arie{Move to Section~\ref{recsys}}

When a work item or an issue is assigned to a developer, the \nalanda{} uses a combination of title and description of the assigned work item as an input search query and provides recommendations about work items or \pr{}s that accomplished similar tasks. Additionally, the system also provides a list of subject matter experts whom a developer can reach out to seek help while working on that work item.

%For example, when a developer is working on a work item about implementing key vault for secret management, a search is performed against the \NalandaDocIndex{} to find the \pr{}s and work items that are relevant based on text similarity, and the people that have expertise with key vault. Then, the results are re-ranked using the socio-technical graph to provide the customized results.

%The \nalanda{} comprises of three primary components: 
%(1) an artifact index, (2) an \expertLower{} index; as well as a search algorithm to traverse these components and provide the recommendations about artifacts and \expertLower{}s. %Details about these components and the search algorithm are explained in the remainder of this section.

\reduceVSpace{}
\subsection{The Nalanda Ranking Algorithm} \label{Algo}
%\arie{Move to Section~\ref{recsys}}

%In this section, we describe the Nalanda ranking algorithm. 
To construct the Nalanda Expert and Artifact recommenders, we devised a ranking algorithm consisting of three steps: 1) querying the artifact index (see Section \ref{docIndex}) to get the relevant \pr{}s and work items, 2) querying the \expertLower{} index (see Section \ref{peopleIndex}) to get a list of relevant \expertLower{}s, 
and 3) re-ranking the results using the Nalanda graph. To that end, we take the following steps:

\textbf{Step 1}: We first construct a query as a combination of the title and description of the work item. Then, we tokenize the query using heuristics that we built for the software engineering domain, such as splitting strings into camel-cased or pascal-cased tokens. We also create n-gram based tokens since we found that bi-grams and tri-grams, such as \texttt{ImapTransfer} and \texttt{MailboxSyncEngine}, capture important information. Next, we filter out stop words \cite{Luhn1960KeyWI}. %We conduct this pre-processing to perform query rewriting and expansion \cite{Liu2020} by generating n-grams.

We employ the BM25 algorithm, which takes care of prioritizing important tokens using Inverse Document Frequency (IDF) scores. It assigns more weight to the documents with a higher overlap with the search query tokens (term frequency), and calculates a relevance score as shown in Equation \ref{eq:bm25}.

% While querying the \NalandaDocIndex{}, a query is constructed from the combination of the work item's title and description. Every \pr{} (and the work item linked to is) is represented as a document. 

\textbf{Step 2}: We query the \expertLower{} index, which contains one document per person. These documents contain the tokens mined from a developer's \pr{}s and work items history. We perform the same pre-processing explained in Step 1 on the query. The BM25 index returns a ranked list of \expertLower{}s.

\textbf{Step 3}: A heuristics-based filtering scheme is used to filter out results with relevance scores below a threshold.
  %  \begin{equation} \label{eq:4}
   %     \Theta = \{n \in \Gamma \mid r(\,n)\, > \omega \}    
    %\end{equation}
%A filter is applied on the relevance score of the set of results returned by the BM25 index. This is done by filtering out the results whose relevance score is less than a threshold to filter out the low-relevance results. 
We determine the threshold empirically to set it at the 75th percentile of the relevance score distribution. 
We determined this value based on a series of experiments to optimize the accuracy of the results while reducing the size of the results set that is passed to the next step.

\textbf{Step 4}: 
We use the Nalanda socio-technical graph to assign proximity scores based on the edge distance between the person performing the search and the results returned by the BM25 index (obtained from Step-3). The proximity score is the length of the shortest path between two nodes. We use the proximity score to re-rank the results. For example, the proximity score is 1 if a \pr{}, work item, or people node is 1-edge away from a developer node, in their shortest path.

\iffalse
We explain this through an example in which we need to provide recommendations for a developer~$D$, given a search query~$Q$. The BM25 index returns a set of entities in response to the search query~$Q$. That is:

\reduceVSpace{}
\begin{equation} \label{eq:2}
    \Gamma = [ PR_1, PR_2, \expertLower{}_1, PR_3, \expertLower{}_2 ]
\end{equation}
PR\textsubscript{1} is the \pr{} that is ranked as the top match based on the relevance score calculated by the BM25 algorithm. Similarly, \expertLower{}\textsubscript{1} is the person ranked as the top expert by the BM25 algorithm. 

The Nalanda graph assigns a proximity score to each entity with respect to the developer performing the search. A new set with proximity scores is constructed. That is,

\reduceVSpace{}
\begin{equation} \label{eq:3}
    \Delta = [ 
        PS^{D}_{pr_1}\,, 
        PS^{D}_{pr_2}\,, 
        PS^{D}_{\expertLower{}_1}\,, 
        PS^{D}_{pr_3}\,, 
        PS^{D}_{\expertLower{}_2}\,
    ]
\end{equation}
PS\textsuperscript{D}\textsubscript{pr\textsubscript{1}}\, is the proximity score between developer node D and  \pr{} node PR\textsubscript{1}. PS\textsuperscript{D}\textsubscript{\expertLower{}\textsubscript{1}}\ is the proximity score between  developer node D and people node \expertLower{}\textsubscript{1}.
\fi

\textbf{Step 5}: We then pick the top-$k$ results from the results set and return them to the user. Items that have a higher BM25 relevance score and are in close proximity to a developer are ranked higher. Furthermore, we use proximity score to break the tie between results where relevance scores are the same.  

%We zip through the sets \textTheta\ and \textPhi\ to calculate the final relevance score for each item in the result set. 
%\begin{equation} \label{eq:5}
%    \Psi = \{i \in \Phi, n \in \Theta \mid n \times 1 / i_n \}    
%\end{equation}

\subsection{Implementation} \label{Impl}
%\arie{Move to Section~\ref{recsys}?}

\begin{figure}
\centering
\includegraphics[width=0.5\columnwidth]{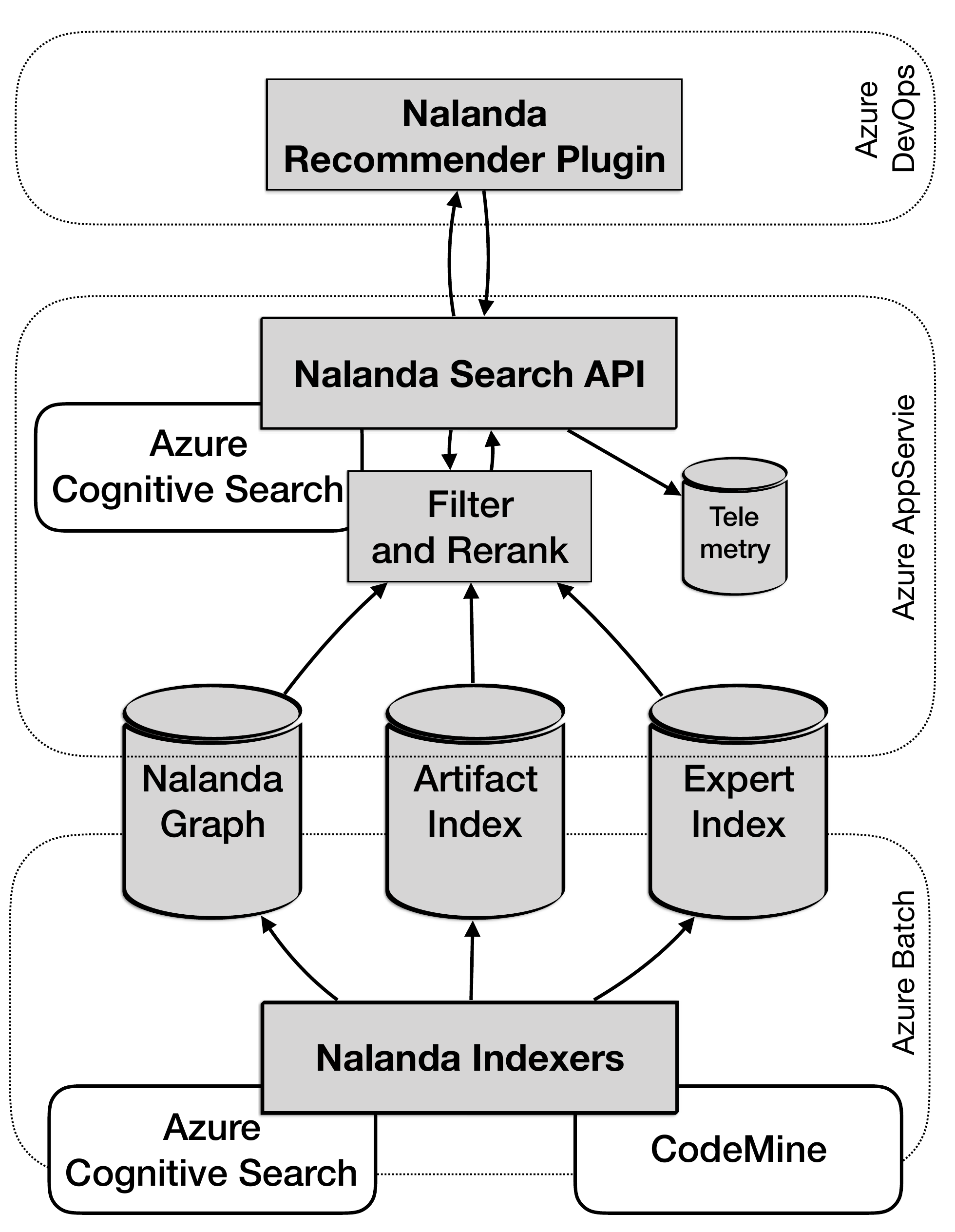}
\caption{\nalanda{} architecture}
\label{SearchArch}
\end{figure}

We implemented the Nalanda search system on the Azure platform, with an emphasis on scalability to thousands of repositories.
The underlying architecture is visualized in Figure~\ref{SearchArch}.

\subsubsection{Nalanda Indexers}

The backbone of the recommendation system is the batch jobs creating and continuously updating the indices and the socio-technical graph, as discussed in Section \ref{design} and shown at the bottom of Figure~\ref{SearchArch}.
%The Nalanda indices and graph require \pr{} and work item data from thousands of source code repositories and need the data to be refreshed periodically.
We leverage CodeMine \cite{6509369} to help us with aggregating source control system data from thousands of repositories.

We rely on Azure Cognitive Search (ACS) \cite{ACS}, which provides BM25 as a service, to store and access the indices. This helps us in alleviating the problems associated with service maintenance, uptime, and scale-out.

%We use Azure Batch \cite{Batch} jobs to run in scheduled intervals to refresh data from both the work item and \pr{} data.

\subsubsection{Search API}

Given the indices and the socio-technical graph, the Nalanda Search API
(shown in the middle of Figure~\ref{SearchArch}) implements the search algorithm from Section~\ref{Algo}.
%Here again we rely on Azure Cognitive Search service (ACS) \cite{ACS} to query the indices.
Each search query (typically initiated by the users in \codebook or through Azure DevOps) passed to the Nalanda search API is processed in real-time. The mean response time for the API call is 1.7 seconds. 

The Nalanda API service asynchronously saves telemetry to an Azure SQL database without impacting the query performance. The telemetry includes the search query, user metadata, search results, click logs, and the API response time. We use this information to evaluate engagement and to improve the performance of the Nalanda search system and the API service.

\begin{table}
\centering
\caption{Evaluation Data Summary}
\label{tab:evaluation-data}
\footnotesize
\begin{tabular}{lp{0.3\linewidth}p{0.3\linewidth}}
 \toprule
 & Artifact Recommendation   & \expert{} Recommendation \\
 \midrule
Index data  &
Pull request title and description, socio-technical graph &
Pull request title and description, socio-technical graph \\
Test set  &
Work Item title and description &
Internal StackOverflow questions \\
Ground truth &
Pull requests linked to the work item &
People who answered the post and people who have answered at least five other questions with the same tag as the post\\
Data set &
80K randomly sampled work items from the \NumRepos{} repositories & 10K randomly sampled questions and answers from StackOverflow\\
\bottomrule
\end{tabular}
\end{table}

\subsection{Quantitative Evaluation} \label{eval}

To understand the efficacy and usefulness of the \nalanda{}, we conduct a large-scale offline evaluation and a user study. %With the offline evaluation, we have two goals: (a) evaluate the accuracy of the \nalanda{} for both \pr{} and \expertLower{} recommendation scenarios, (b) conduct a comparative study to find the optimal scheme for both the artifact and the expert indexes. For the comparative study, we do additive experiments where we evaluate improvement in performance after adding properties to the indexes. 

\subsubsection{Experiment Setup}
We randomly sample 80,000 work items from the \NumRepos{} repositories such that there are at least 10 work items selected from each repository. Subsequently, we use the title and description of each of these work items as the input to the Nalanda search API. We expect the right \pr{}s and people to be returned from the search API. 

Since a recommendation system like ours does not exist in the company, we do not have a ground truth to conduct a large-scale evaluation. Therefore, we rely on the \pr{}s manually tagged by developers to the work items in \ado{} to build the evaluation dataset. To create the ground truth dataset for the expert recommendations, we leverage the private instance of StackOverflow deployed at \companyX{}. Details about the index, ground truth, and test sets used for these experiments are shown in Table~\ref{tab:evaluation-data}.

%We use the title of the StackOverflow questions as the input passed to the search API. we use a combination of (a) the set of people who answered the question, and (b) set of people who answered at least five other questions tagged with at least one of the tag(s) as the given question.
\subsubsection{Results} We use two commonly used metrics for recommender systems: 1. Top K accuracy, which measures the number of times the correct item is found in  the top K recommendations 2. Mean Reciprocal Rank (MRR), which calculates the reciprocal of the rank at which the first relevant document was retrieved \cite{Craswell2009}.

\begin{table}
\caption{Evaluation and Comparative study for Artifact Recommendation}
%\begin{adjustwidth}{-10cm}{-10cm} % To fix long width table alignment %
\centering
\label{tab:pr-ablation}
\scriptsize
\begin{tabularx}{\columnwidth}{l|r|r|r|r|r|r|}
~                                                                    & \multicolumn{2}{c|}{\textbf{K = 3}} & \multicolumn{2}{c|}{\textbf{K = 5}} & \multicolumn{2}{c|}{\textbf{K = 10}}  \\ 
\cline{2-7}
\textbf{Indexed Properties}                                            & \textbf{Accuracy} & \textbf{MRR}    & \textbf{Acc} & \textbf{MRR}    & \textbf{Acc} & \textbf{MRR}      \\ 
\cline{2-7}
PR metadata                                                        & %0.2647            & 0.2324          & 0.2868            & 0.2769          & 0.3191            & 0.3031            \\
0.26            & 0.23          & 0.29            & 0.28          & 0.32            & 0.30            \\
PR attributes                                                        &             &           &             &           &             &             \\
~ +  PR title                                                        & %0.38              & 0.3617          & 0.434             & 0.4142          & 0.5111            & 0.4806            \\
0.38              & 0.36          & 0.43             & 0.41          & 0.51            & 0.48            \\

~ +  PR description                                                         %& 0.494             & 0.4714          & 0.538             & 0.5066          & 0.5962            & 0.5871            \\
   & 0.49             & 0.47          & 0.53             & 0.51          & 0.60            & 0.59            \\
\begin{tabular}[c]{@{}l@{}}~ + socio-technical graph~\end{tabular} & %\textbf{0.718}   & \textbf{0.7092} & \textbf{0.74 }    & \textbf{0.7267} & \textbf{0.7778}   & \textbf{0.7674}  
\textbf{0.71}   & \textbf{0.71} & \textbf{0.74}    & \textbf{0.73} & \textbf{0.78}   & \textbf{0.77}
\end{tabularx}
%\end{adjustwidth}
\end{table}

\begin{table}
\caption{Evaluation and Comparative study for Expert Recommendation}
%\begin{adjustwidth}{-5cm}{-5cm} % To fix long width table alignment %
\centering
\label{tab:people-evaluation}
\scriptsize
\begin{tabularx}{\columnwidth}{l|r|r|r|r|r|r|}
~                           & \multicolumn{2}{c|}{\textbf{K = 3}} & \multicolumn{2}{c|}{\textbf{K = 5}} & \multicolumn{2}{c|}{\textbf{K = 10}}  \\ 
\cline{2-7}
\textbf{Indexed Properties} & \textbf{Accuracy} & \textbf{MRR}    & \textbf{Acc} & \textbf{MRR}    & \textbf{Acc} & \textbf{MRR}      \\ 
\cline{2-7}
%\pr{} metadata                    & 0.3506            & 0.291           & 0.3855            & 0.3349          & 0.4251            & 0.3735          \\
PR metadata                    & 0.35            & 0.30           & 0.39            & 0.33          & 0.43            & 0.38            \\

PR attributes                  &                   &                 &                   &                 &                   &                   \\
%~  +  \pr{} title             & 0.5150            & 0.456           & 0.5416            & 0.4835          & 0.587             & 0.5277            \\
~  +  PR title             & 0.51            & 0.46           & 0.54            & 0.49          & 0.59             & 0.53            \\
%~ +  \pr{} description       & 0.6001            & 0.5351          & 0.6435            & 0.587           & 0.6825            & 0.6105            \\
~ +  PR description       & 0.60            & 0.54          & 0.64            & 0.59           & 0.69            & 0.61            \\
%~ + socio-technical graph   & \textbf{0.6303}   & \textbf{0.5901} & \textbf{0.6811}   & \textbf{0.632}  & \textbf{0.7455}   & \textbf{0.67}    

~ + socio-technical graph   & \textbf{0.63}   & \textbf{0.60} & \textbf{0.69}   & \textbf{0.63}  & \textbf{0.75}   & \textbf{0.67}
\end{tabularx}
%\end{adjustwidth}
\end{table}

Table \ref{tab:pr-ablation} shows the results from the evaluation for different values of K. We can see that incorporating more attributes of the \pr{}, such as its title and description, improves both the MRR and accuracy considerably. Furthermore, re-ranking the results using the Nalanda graph also substantially improves the recommendations. Similar improvements can be noticed for the expert recommendation task too (Table \ref{tab:people-evaluation}).
%For $K=10$, the accuracy is increased from $0.60$ to $0.78$ $(+30.45\%)$,
%whereas the MRR is improved from $0.59$ to $0.77$ $(+30.71\%)$ after the re-ranking. 
% 0.6 to 0.78; 0.59 to 0.77

%We show the results for the \expertLower{} recommendation task in Table \ref{tab:people-evaluation}. For the baseline setup with just the pull request meta-data, the accuracy ranges from 0.35 (for K = 3) to 0.43 (for K = 10). Both the accuracy and MRR increase substantially as we add \pr{} titles and descriptions to the index. Further, reranking using the graph improves accuracy from 0.60 to 0.63 (+5\%) for K=3 and from 0.69 to 0.75 (+8.7\%) for K=10.

\subsection{User Perception}\label{userstudy}

We conducted a user study among developers regarding the usefulness of the recommendations. We selected ten participants (identified as P1--P10) from \companyX{} to evaluate the recommendations on their recently completed work items. %The participants are the owners of ten randomly chosen work items that were closed within the last one month from the date of study. %As our system is still in beta testing phase, we cannot pass recommendations on newly created work items directly.
%Instead, we leverage recently completed work items to conduct our user study. Our subjects are likely to have a good recollection of the context of these work items as they worked on them recently.

\subsubsection{Participants and Protocol} 
\label{sec:protocol}
We conducted semi-structured interviews, which were conducted remotely, and ranged from 15-30 minutes. The average experience of the subjects is 7.7 years in the company and ranged from 10 months to 21 years. %Six of the ten participants have a tenure less than or equal to three years in the same team and four of them have greater than three years of experience. Seven of the ten participants are men and three are women. Four participants are from an offshore development centre and six work in the headquarters.

We employ a one-group pretest-posttest pre-experimental design~\cite{Campbell1963ExperimentalAQ}. %This type of experiment is called pre-experimental to indicate that it does not meet the scientific standards of experimental design \cite{Babbie1969ThePO}, yet it allows us to report on facts of real user behavior, even those observed in under controlled, limited sample experiences. The primary purpose of such test is to understand if there are differences in the subjects' expectation versus perception of the recommendation system. 
We used the Likert scale \cite{Robinson2014} for rating the responses. The respondents can provide their responses on a 1 to 5 scale, ranging from `strongly disagree' to `strongly agree'. We posed them the questions listed below.

\emph{When you are working on a work item, how useful would the recommendations be in completing the work item (on a scale of 1 to 5):}

\begin{enumerate}
    \item You are going to refer to the work items and \pr{}s recommended as inspiration and informal documentation on accomplishing the work item. 
    \item You would likely reach out to the recommended people for consultation on accomplishing the work item.
\end{enumerate}

\begin{figure}
\centering
\includegraphics[width=1\columnwidth]{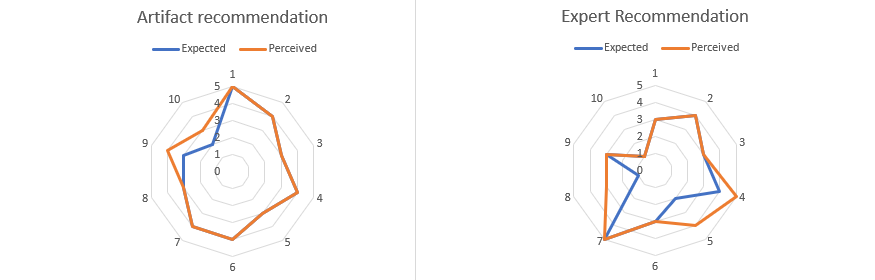}
\caption{Participants' evaluation of the Nalanda system}
\label{prepost-radar}
\reduceVSpace{}
\end{figure}

\subsubsection{Results}
\textbf{Do the users feel work item, \pr{}, and \expertLower{} recommendations are useful?}
The participants expressed that the Nalanda recommendation system can be a great value addition to the software development process. 60\% of the participants rated `agree' or `strongly agree' when asked whether they find the artifact recommendations useful and 40\% responded favorably to the \expertLower{} recommendations.

Through question 2 we measure the difference between the expectations the participants had of a hypothetical recommendation system with the \nalanda{}. This question measures the dependent variable for the user study (introduction of the Nalanda system). In Figure \ref{prepost-radar}, the radar chart shows some differences between the participants' original expectations and their perception of the Nalanda recommendations.

\begin{figure}
\centering
\includegraphics[width=0.75\columnwidth]{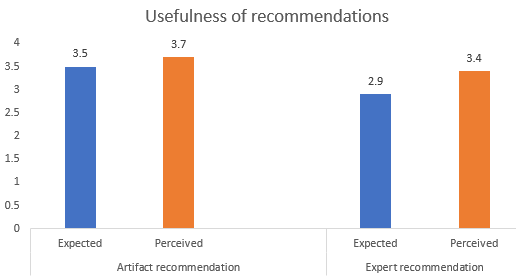}
\caption{Expectations and perceptions of the Nalanda system}
\label{pre-post-usefulness}
\reduceVSpace{}
\end{figure}

 These differences can be observed better in Figure \ref{pre-post-usefulness}, in which the averages of the rating are shown. The difference in expectation versus perception was more apparent with the \expertLower{} recommendations compared to the artifact recommendations.

 To offer an impression, we list some typical quotes (positive and negative) that we received from the developers.

\begin{displayquote}
    \textit{\quotes{A tool like this will help greatly to understand the processes involved in pushing my changes through.}}
\end{displayquote}

\begin{displayquote}
    \textit{\quotes{It is great to see the recommendations about people to talk to. My team is large, mostly remote, and I am new. So this is very helpful.}}
\end{displayquote}

\begin{displayquote}
    \textit{\quotes{Finding people to talk to has never been a problem for me as I have been working in the same org for a while.}}
\end{displayquote}

%% file: Discussion.tex
\section{Discussion} \label{Discussion}

\subsection{Outlook}
In the future, we anticipate both the \NalandaGraphPlat{} and \codebook{} to be scaled out significantly inside Microsoft in terms of the number of repositories and users. %Looking at the success, favorability, and the scale of the deployment of the \NalandaGraphPlat{} and \codebook{} inside Microsoft, we believe the systems and the techniques has applicability beyond Microsoft. 
Furthermore, we see opportunities for implementing the \nlg{} platform as a service on top of the open-source software development data mined from systems like GitHub. 

We also expect the \nalanda{} to be employed at substantially more software development systems at \companyX{}. %From the evaluation results and the responses from the user study we are optimistic about \nalanda{}'s reach and impact. 
%Similar to \codebook{} we see opportunities to implement the \nalanda{} in open source systems such as GitHub. %Future research surrounding the \nalanda{} might entail improving its accuracy by learning from user feedback and interaction models or by leveraging more \pr{} attributes without sacrificing scalability. Beyond artifact and \expertLower{} recommendations, 
Future research could entail including other types of useful recommendations such as internal and external documentation, tutorials, and recommendations from question-answer forums.

The rich socio-technical data in the \nlg{} in combination with advances in deep learning and Graph Convolutional Networks (GCNs), hold promise for applications such as neural reviewer recommendations by leveraging the socio-technical structures. The \NalandaGraphPlat{} lays the foundation to bring various techniques from the graph representation world \cite{GraphTechniques}, such as link prediction, social network analysis, etc. into the software engineering and analytics domain.  

\subsection{Threats and Limitations} 

\subsubsection{Internal Validity}
Conducting trustworthy  experiments on data collected from thousands of repositories is challenging
especially due to the problems of avoiding data leakage and obtaining credible ground truth. In our experiments, we addressed this (see Table~\ref{tab:evaluation-data}). While our results are highly promising and an important first step, more experiments are needed to better
understand  the true nature of the graph's contribution to expert and artifact recommendation.

The risk of response bias is minimal in all our studies because all the participants of the user study are organizationally distant from the people involved in building this system. However, there remains a small chance that people in the user study may be positive about the system because they want to make the developers who are from the same company motivated and happy.

\subsubsection{External Validity}
The context of our evaluation and user studies is a software development company with a large number of developers. These developers work on a portfolio of products across many contexts and domains. By conducting a user study within one single company, we were able to control for factors like culture, tooling, frameworks, and programming languages. However, our results may not be generalizable across all developers in all contexts. Hence, our results are not verified in the context of other organizations or the open-source community. Therefore, our findings may be limited and warrant further research. Future work could investigate user interfaces, integrating our findings with design guidelines that span usability and technical and organizational complexity \cite{jaferian2008guidelines}.

%% file: RelatedWork.tex
\section{Related work}\label{related}

%Related work falls into the following areas: graphs representations of software engineering activities, social networking in software engineering, and newsfeeds in software engineering.

\paragraph{Graph Representations}
Hipikat \cite{Hipikat} is one of the earliest works to build a graph for software development entities like tasks, file versions, and documents, using a fixed schema. It was built for onboarding new hires quickly by providing easy access to relevant artifacts. \originalcodebook \cite{begel2010codebook} builds a prototype graph consisting of various software development entities. The data was mined from software repositories for a single team at Microsoft. Bhattacharya et al. \cite{BhattacharyaICSE2012SeverityPrediction} use graph representation of source code and bug tracking information to construct  predictors for software engineering metrics like bug severity and maintenance efforts. Other applications of graph representations of software artifacts include visualizing relationships among project entities \cite{Tesseract2009} and extracting changes in variability models \cite{Fever2018}. Compared to all this work, the scale of the \nlg{} is significantly larger with \NumNodesM{} nodes and \NumEdgesM{} edges, and Nalanda has been designed to be actively used in a production setting.

\paragraph{Source Control Dashboards}
GitHub offers a dashboard for developers on its homepage \cite{GitHubDashBoard}. It displays the repositories, active pull requests, and issues (work items). A key limitation of the GitHub dashboard is there exists no notion of a ``Team feed'' in Github unlike \codebook. Team feed helps the discovery of the items worked on by other team members. 

\paragraph{Information Needs}
Ko et al.~\cite{ko2007information} studied information needs in colocated development teams. 
Fritz and Murphy \cite{Fritz2010InfoFragments} provide a list of questions developers ask for the most frequently sought-after information within a project. 
Information needs have also been studied in the context of change tasks~\cite{sillito2008asking}, inter-team coordination~\cite{begel2010codebook} and software analytics~\cite{buse2012information, begel2014analyze, huijgens2020questions}
Through its applications, Nalanda can efficiently address most information needs related to people, code, and work items. For example, the answer to questions such as ``Who is working on what'' and ``What are coworkers working on right now'' is easily available in the \codebook application.

\paragraph{Artifact Recommendation systems for software developers}
Recommendation systems for software engineering aim at assisting developers with activities such as code reusability, writing effective bug reports, etc. \cite{5235134}. Tools like CodeBroker \cite{10.1145/581339.581402} help in finding the relevant code samples extracted from the standard Java documentation generated by Javadoc from Java source programs and deliver the suggestion to the Emacs editor. Anvik et al. \cite{msr1} proposed a semi-automated method to assign bug reports to reporters based on their expertise using a machine learning algorithm. Mockus and Herbsleb\cite{mockus2002expertise} used quantity as a measure of expertise. 
%The tool is presented in the form of HTML pages with search engine and hierarchical navigation. 
Fu et al.\cite{msr2} used the node2vec algorithm to convert file entities within projects into knowledge mappings. They proposed four features to capture the social relationships between developers. Devrec\cite{msr3}, a developer recommendation system, mines the development activities of developers in GitHub and StackOverflow to recommend collaborators for a given project. Hammad et al \cite{msr5} use keywords from the textual content of commits. %Each expert is associated with the extracted keywords for a recommendation. 
On the other hand, Canfora et al \cite{msr6} use mailing lists and versioning systems to recommend experts for newcomers joining a software project. Compared to these approaches, the \nalanda{} is designed to be highly scalable and provide responses in real-time. %As the integration happens at Azure DevOps level, developers will have the information that they'd need to get started on a work item, within the same work item itself.

%% file: Conclusion.tex
\reduceVSpace{}
\section{Conclusion}
In this paper, we seek to build a \emph{large scale} software analytics data platform named Nalanda with two subsystems (the \NalandaGraphPlat{}and the \NalandaIndexPlat{}). The \NalandaGraphPlat{} consists of a socio-technical graph encompassing the entities, people, and relationships involved in the software development life cycle. The \NalandaIndexPlat{} is an \emph{enterprise scale} index system that can be used to support a wide range of software engineering tasks such as recommendation, and search.

We built the \NalandaGraphPlat{} using software development activity data from \NumRepos{} source code repositories. The graph consists of \NumNodes{} nodes and \NumEdges{} edges. To the best of our knowledge, it is the largest socio-technical graph built to date using private software development data. Similarly, the \NalandaIndexPlat{} contains 8,018,320 documents in its artifact index and 61,428 documents in its \expertLower{} index with data ingested from \NumRepos{} repositories at \companyX{}.

The Nalanda platform and its applications (\codebook{} and \nalanda{}) help in developing awareness of each other's work, and building connections between developers across repositories, while offering mechanisms to discover information while managing information overload. We also seek to address the problems of information discovery by finding related work items and experts for software developers.  

Based on organic growth alone, \codebook{} has Daily Active Users (DAU) of 290 and Monthly Active Users (MAU) of 590.  A preliminary user study shows that 74\% of developers and engineering managers surveyed are favorable toward continued use of \codebook{} for information discovery. The \nalanda{}, with the help of the socio-technical graph for customization, lifted the accuracy of artifact recommendations by 30.45 percentage points to 0.78. In a study with ten professional software developers, participants agreed that a system like \nalanda{} could reduce the time spent and the number of places needed to visit to find information.

In the future, we anticipate both the \NalandaGraphPlat{} and \codebook{} to be scaled out significantly inside Microsoft in terms of the number of repositories and users. We believe the systems and the techniques have applicability beyond Microsoft. Furthermore, we see opportunities for implementing the \nlg{} platform as a service on top of the  open-source data mined from platforms like GitHub. 